# Ultra-Strong Coupling Effects with Quantum Metamaterials.

Jonathan Plumridge, Edmund Clarke, Ray Murray and Chris Phillips, *Experimental Solid State Group, Physics Department, Imperial College, Prince Consort Roa,d London SW7 2AZ, UK.*

We study a semiconductor-based quantum metamaterial which has the optical characteristics of a metal in two directions, but behaves like a collection of artificial atoms, whose properties can be designed-in using quantum theory, in the third. We find that it supports a new type of guided collective plasma resonance (CPR) mode which exhibits efficient optical coupling and long propagation distances. Furthermore, the coupling of the CPR mode with the 'artificial atom' transition leads to a case of "Ultra-Strong-Coupling", demonstrated by a large vacuum Rabi splitting of 65 meV, a sizable fraction (21%), of the bare intersubband energy.

Surface-plasmon-polaritons (SPP's) have long been known to be supported at the planar boundary between two materials whose dielectric constants are opposite in sign, and their strongly enhanced fields have proven valuable for the realisation of subwavelength optics[1], non-linear optics[2], optical sensors[3], and mid-IR diode laser cavities[4].

Normally the negative dielectric is a strongly absorbing metal. This usually leads to heavily damped modes which propagate for < 100 μm[1], but recently it has been shown that thin metal slabs can support propagating guided plasmon modes[5] known as "Long Range Plasmon" (LRP) modes, which propagate over mm distances[5].

Our experiments use semiconductors. Their mature fabrication technology lets us structure samples microscopically, with extremely smooth interfaces, to support analogues of the metallic LRP modes. However, we can also structure them on a nanoscopic scale, which is both small enough to be invisible to the electromagnetic (EM) fields (< 1/100 of a wavelength), and small enough to modify, through the rules of quantum mechanics, the behaviour of the electrons in the slab. EM fields now see a slab whose electrons are free to move in the x-y slab plane, giving a metallic response to light polarised in that direction, but whose z-motion is completely quantised, so light sees a single atom-like Lorentzian oscillator response corresponding to single electron intersubband transitions (ISBT's) between the confined states of the quantum wells.

The approach can be viewed as an extension of the metamaterial concept[6]. With "Quantum Metamaterials", we exploit not only the wavelike nature of EM radiation but also we exploit the wavelike nature of matter; we control the electrons' behaviour at the quantum level and instead of using just Maxwell's equations to design-in the resonances, we also use Schroedinger's [7].

At a given MQW doping level the plasma frequency, $\omega_p$, determines the frequency at which the in-plane dielectric constant flips from negative to positive. The critical point to note though is that when the MQW slab is thin compared with the optical wavelength, mode volumes, dispersion curves and propagation distances are completely unaffected by this sign change[7]. However, because the E- field in the thin slab is only in the z-direction, we can use the quantum metamaterial approach to design out the losses, enabling the CPR mode to propagate for centimetres[7].

The ISBT energy is dispersionless and is governed by the nanoscopic QW thicknesses (its oscillator strength is determined by the QW doping). On the other hand, the dispersion characteristics of the CPR mode allows it to be tuned through the ISBT energy in angle resolved spectroscopic measurements.

Demonstrations of Strong-coupling (SC) typically involve a dipole-allowed electronic transition coupled to the photon modes of a small optical cavity[8,9]. If the two oscillators are sufficiently localised in energy (i.e. a sharp transition line coupled to a high finesse cavity) and space (i.e. a small optical mode volume) their Vacuum Rabi coupling energy, $\hbar\Omega_{VR}$, may exceed the linewidths of both the electronic transition and the cavity resonance. Tuning the modes' energies through degeneracy now generates anticrossing behaviour and mixed light-matter modes[10]. However, the condition of Ultra-strong-coupling occurs when the magnitude of the anticrossing, $\hbar\Omega_{VR}$, becomes comparable to the energy of the bare electronic transition. Under these circumstances the effects of quantum electrodynamics start to play a significant role, and may be exploited for potential applications such as the formation of correlated photon pairs[11]. Here we study anticrossing behaviour to show that the Ultra-Strong-Coupling regime (USC) can be realised in a robust and manufacturable solid-state system, even though the linewidths are large by the standards of previous SC experiments.

The multiple-quantum-well (MQW) sample is grown by molecular-beam-epitaxy and comprises 50 periods, each with a 6nm wide GaAs QW separated by a 30 nm $Al_{0.35}Ga_{0.65}As$ barrier. The n-doping ($n_s \sim 6.5 \times 10^{11}$ cm$^{-2}$ per well) populates only the lowest confined QW state, giving a single ISBT resonance. Electrons in neighbouring wells are quantum-mechanically isolated, but they couple via the Coulomb interaction. The MQW slab is clad above and below with the same undoped $Al_{0.9}Ga_{0.1}As$ material; both layers are thick enough (> 2μm) to closely approximate the LRP ideal of perfectly symmetric semi-infinite cladding, by keeping the CPR mode away from the GaAs substrate and the 350nm thick capping layer.

The wafer was cut and polished into a rhombohedral prism (fig.1) for polarised, angle-resolved reflectivity spectra to be taken, at an angular resolution of Δφ~ 2.5°. Residual substrate absorption makes it impossible to get good quantitative s- and p- polarised spectra in this situation, but the ratio of p- to s- polarised reflectivity can be measured and calibrated with high confidence.

Tuning the internal angle of incidence, θ, to the slab, changes the in-plane wavevector $k_x = |\underline{K}| \sin(\theta) [\varepsilon_{GaAs}]^{1/2}$, where $\underline{K}$ is the free-space wavevector of the beam outside the sample, and $\varepsilon_{GaAs}$ =10.36 the dielectric constant of the GaAs substrate[12]. This scans through the CPR dispersion curve, and allows the two excitations to be tuned through

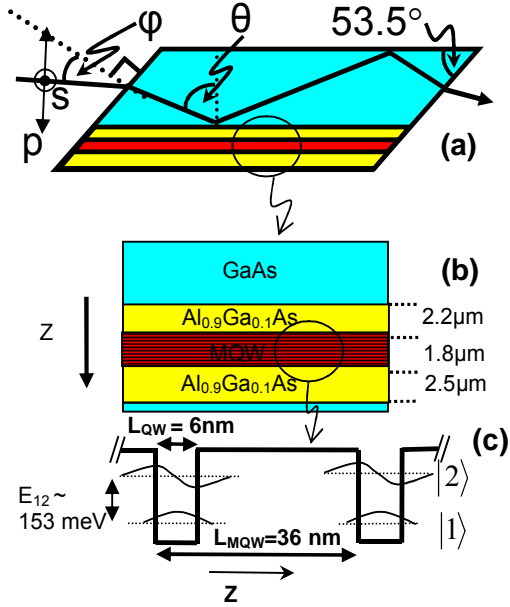

FIG. 1 a) Prism coupling scheme for mapping out the dispersion curves of the mixed Collective Plasma Resonance [CPR] and Intersubband Transition ([ISBT] modes. Internal angles, θ, are related to the external ones, φ, through $\theta = 53.5° + \sin^{-1}[\sin(\varphi)/\sqrt{\varepsilon_{GaAs}}]$.(b) Detail of the microscopic slab structure designed to support the CPR mode.(c) Detail of the nanoscopic "artificial atom" structure comprising a 50-period stack of doped quantum wells.

degeneracy.

The measured spectra (Fig.2) each show two pronounced features, an ISBT-like excitation at ~1200 cm$^{-1}$, and a CPR like feature, which angle tunes from ~2400cm$^{-1}$ to ~1400 cm$^{-1}$. The ISBT red shifts, by up to ~200 cm$^{-1}$ /24meV, as the CPR energy is tuned towards it, and clear anti-crossing behaviour appears, with an anticrossing energy, $\hbar\Omega_{VR}$ ~ 65 meV, some 8 times larger than the linewidth of either excitation.

We model these curves with an optical transfer matrix (TM) method[13]. The sub-wavelength nature of the nanoscale structuring fully justifies an effective-medium approach[12], and we write the MQW's optical response as a complex, anisotropic dielectric tensor

$$\bar{\varepsilon}_2 = \begin{pmatrix} \varepsilon_{xx} & 0 & 0 \\ 0 & \varepsilon_{yy} & 0 \\ 0 & & \varepsilon_{zz} \end{pmatrix} \quad (1)$$

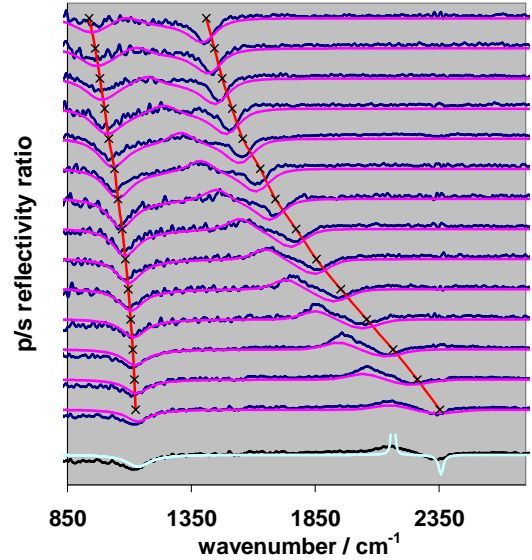

FIG.2 Ratio of p- to s- polarised angle resolved reflectivity spectra at external angles ranging from φ= 61° to φ= 35° (highest trace), in 2° steps Successive traces shifted vertically, by an amount corresponding to a 100% dip in reflectivity, for clarity. Blue curves: experimental data, showing ISBT at 1100 cm$^{-1}$ and the CPR, at ~2300 cm$^{-1}$. Light blue curves; model reflectivity calculation for the p- to s- polarised reflectivity ratio, including broadening effects arising from the limited angular resolution. Lowest trace; φ= 61° modelled spectrum without broadening, showing the high-Q low loss nature of the CRP (the truncated positive excursion at ~2070 cm$^{-1}$ is an artefact from the s- polarised reflectivity nearing zero in the model). Red curves, guides to the eye showing the positions (crosses) of the experimental p-polarised reflectivity minima.

$$\varepsilon_{xx} = \varepsilon_{yy} = \varepsilon_y - \left(\frac{\omega_p}{\omega}\right)^2 \frac{1}{1 + i\left(\frac{\gamma_1}{\omega}\right)} \quad (2)$$

$$\frac{1}{\varepsilon_{zz}} = \frac{1}{\varepsilon_z} - \left(\frac{\omega_p^2 f_{12}}{2\omega\gamma_2\varepsilon_w}\right) \Big/ \left[\left(\frac{E_{12}^2 - \hbar^2\omega^2}{2\hbar^2\gamma_2\omega}\right) - i\right]. \quad (3)$$

Where $\varepsilon_y$ and $\varepsilon_z$ are the mean effective background dielectric constants, given by $\varepsilon_y = (1-L_{qw}/L_{mqw})\varepsilon_b + (L_{qw}/L_{mqw})\varepsilon_w$ and $\varepsilon_z^{-1} = (1-L_{qw}/L_{mqw})/\varepsilon_b + (L_{qw}/L_{mqw})/\varepsilon_w$, where $L_{qw}$ and $L_{mqw}$ are the QW width and MWQ period respectively, and $\varepsilon_b(\varepsilon_w)$ are the background dielectric constants corresponding to undoped barrier (well) materials.

The in-plane electrons' response (eq.2) is Drude-like with a plasma frequency $\omega_p = [n_s e^2/ m^* \varepsilon_b \varepsilon_z L_{mqw}]^{1/2}$, where $n_s$ is the areal electron density per well, $m^*$ the electron effective mass and the other symbols have their usual meanings. The out of plane electrons' response (eq.3) is that of a single ISBT "artificial atom" Lorentzian oscillator, at energy $E_{12}$, with an oscillator strength $f_{12} = 2m^* E_{12} |z_{12}|^2 / \hbar^2$. $z_{12} = \langle 1|z|2\rangle$ is the intersubband dipole matrix element and $\gamma_2$ the electron dephasing rate. Although each individual QW well is small compared with optical wavelengths $[L_{mqw} << 2\pi c\, (\varepsilon_{zz})^{-1/2}]$ the overall thickness, $d= N*L_{mqw}$, of the slab containing N wells, need not be. We use $\varepsilon_b$=9.88 and $\varepsilon_w$=10.36 and $\varepsilon_1$ = 8.2 for the dielectric constants of $Al_{0.35}Ga_{0.65}As$, GaAs and $Al_{0.9}Ga_{0.1}As$ respectively[12].

We use $z_{12}$ = 1.9 nm (from separate wave function calculations), $m^*$ =0.067 $m_0$ (where $m_o$ is the free electron mass) and $E_{12}$ = 153meV. The ISBT linewidth, measured as 7.5 meV in a separate absorption experiment, gives $\gamma_2 \sim$ 1.1 x $10^{13}$ sec$^{-1}$, and in the absence of more detailed knowledge of the in-plane scattering mechanisms, we set $\gamma_1 = \gamma_2$.

In terms of excitation energies and integrated absorption strengths the agreement between experiment and theory (Fig 2) is excellent, and both generate anti-crossing energies (~65meV) some ~8.5 times larger than the ISBT linewidths, i.e. $\hbar\Omega_{VR}/E_{12}$ = 21% and into the USC regime. At high angles of incidence the strong angle-dependence of the CPR energy means the measured CPR linewidths are limited by the $\Delta\varphi \sim 2.5°$ angular resolution. The broad positive excursions, just to the red of the highest energy CPR modes are due absorption from a damped s-polarised mode, which does not couple to the ISBT. This assignment can be confirmed by analysing the field profiles at those energies in the TM model, and by an independent analytical treatment[7]. With this present, non-optimised sample the TM model predicts propagation distances of 2.1mm and 1.34 mm at photon energies (wavelengths) of 88meV (14μm) and 315 meV (3.94 μm) respectively.

Using the dimensions of the experimental sample we find that the TM model accurately reproduces the $n_s^{\frac{1}{2}}$ splitting energy predicted in reference 8, and shows that $\hbar\Omega_{VR}$ values in excess of 90meV are possible at realistic doping levels of $n_s \sim 10^{12}$ cm$^{-2}$ (where we predict $\hbar\Omega_{VR}/E_{12}$ to be 58%, well into the USC regime). Compared with recent microcavity samples[14] using a single-side surface plasmon waveguide ($\hbar\Omega_{VR}$ = 42 meV and $\hbar\Omega_{VR}/E_{12}$ = 17%), and which use the same overall MQW doping levels as here, we see the CPR resonance effect has increased the coupling energy by 54%. This factor arises from the higher Q (lower losses) achievable with a cavity design based upon CPR resonances.

As well as being capable of being excited and probed optically, the CRP mode offers the potential for being rapidly switched, in devices where the electron density is modulated by standard gating methods. Such a method could well be used to produce correlated photon pairs[11]. Stepped QW potential profiles could also be incorporated which would allow the ISBT energy to be Stark-tuned in and out of the USC regime[15], so a single CPR excitation could be made to interact with separate groups of artificial atoms as it is guided through a patterned device. As a consequence, these CPR guided modes have the potential to form the basis for a new range of compact and coherent opto-electronic switches and circuits.

The authors would like to thank Paul Stavrinou for stimulating discussions. Funding from the UK Engineering and Physical Sciences Research Council is gratefully acknowledged.